\documentclass[journal]{IEEEtran}

\usepackage{booktabs}
\usepackage{graphicx}
\usepackage{multirow}
\usepackage{adjustbox}
\usepackage{float}
\usepackage{makecell}
\usepackage{cite}
\usepackage{amsmath,amssymb,amsfonts}
\usepackage{algorithmic}
\usepackage{textcomp}
\usepackage{xcolor}
\usepackage{stfloats} 
\usepackage{tabularx}
\usepackage{diagbox}
\usepackage{enumitem}

\begin{document}

\title{Unified Architecture and Unsupervised Speech Disentanglement for Speaker Embedding-Free Enrollment in Personalized Speech Enhancement}

\author{Ziling~Huang,~\IEEEmembership{Student Member,~IEEE,} Haixin~Guan,~\IEEEmembership{Member,~IEEE,}
        Yanhua~Long,~\IEEEmembership{Member,~IEEE}

\thanks{Yanhua Long is the Corresponding author. Ziling Huang, Yanhua Long are with Shanghai Engineering Research Center of Intelligent Education and Bigdata, 
Shanghai Normal University, Shanghai, 200234, China. Yanhua Long is also with the SHNU-Unisound Natural Human-Computer 
Interaction Lab, Shanghai Normal University. (e-mail: hzlkycg111@163.com; yanhua@shnu.edu.cn). 
Haixin Guan are with the Unisound AI Technology Co., Ltd., Beijing, China (e-mail: guanhaixin@unisound.com).}}

\maketitle

\begin{abstract}

Conventional speech enhancement (SE) aims to improve speech perception and intelligibility by 
suppressing noise and reverberation without requiring enrollment speech, whereas personalized speech 
enhancement (PSE) addresses the cocktail party problem by extracting a target speaker’s speech using 
enrollment speech as a reference. While these two tasks tackle different yet complementary challenges 
in speech signal processing, they often share similar model architectures, with PSE incorporating an 
additional branch to process enrollment speech. This suggests the possibility of developing a unified 
model capable of efficiently handling both SE and PSE tasks, thereby simplifying deployment while maintaining 
high performance. However, PSE performance is highly sensitive to variations in enrollment speech, 
such as differences in emotional tone, duration, and semantic content, which limiting its robustness 
in real-world applications. 
To address these challenges, we propose two novel models, USEF-PNet and DSEF-PNet, both extending our 
previous SEF-PNet framework. USEF-PNet introduces a unified architecture for processing enrollment speech, 
integrating SE and PSE into a single framework to enhance performance and streamline deployment. 
Meanwhile, DSEF-PNet incorporates an unsupervised speech disentanglement approach by pairing a mixture 
speech with two different enrollment utterances and enforcing consistency in the extracted target speech. 
This strategy effectively isolates high-quality speaker identity information from enrollment speech, 
reducing interference from factors such as emotion and content, thereby improving PSE robustness. 
Additionally, we explore a long-short enrollment pairing (LSEP) strategy to examine the impact of 
enrollment speech duration during both training and evaluation. Extensive experiments on the Libri2Mix and 
VoiceBank-DEMAND datasets demonstrate that our proposed USEF-PNet, DSEF-PNet all achieve substantial performance improvements, with random enrollment duration performing slightly better. Our source code, model checkpoints, and datasets will be publicly available at https://github.com/isHuangZiling/UDSEF-PNet. 

\end{abstract}


\section{Introduction}

\IEEEPARstart{I}{n} real-world applications such as smart home devices and multi-speaker conference systems, 
managing overlapping speech, environmental noise, and room reverberation poses significant challenges. 
A well-known problem in multi-speaker scenarios is the ``cocktail party problem” \cite{cherry,haykin}, 
wherein humans can effortlessly focus on a specific speaker and shift attention, yet replicating this 
capability in machines remains a significant challenge. 
Two effective approaches to address these challenges are conventional speech enhancement (SE) and 
personalized speech enhancement (PSE). SE focuses on removing noise and reverberation to improve speech perception and 
intelligibility, with substantial progress driven by deep learning techniques 
\cite{segan,metricgan,mp-senet,real-time,dual-branch,dpt-fsnet,phasen}, while 
PSE builds upon SE techniques by requiring enrollment speech to extract  
and enhance the target speaker's speech in multi-speaker noisy environments \cite{pseoverview1,pseoverview2}. 

Many state-of-the-art models for PSE and SE share similar architectures. 
For instance, successful SE models such as DPCCN \cite{sdpccn}, BSRNN \cite{pbsrnn}, Sepformer \cite{x-sepformer}, 
and DeepFilterNet \cite{deepfilternet} have their corresponding PSE adaptations (e.g., sDPCCN, pBSRNN, X-Sepformer, and 
pDeepFilterNet), which integrate additional branches to handle enrollment speech. However, simply 
because PSE requires an additional branch to process enrollment speech while sharing the same 
overall model architecture, training two separate sets of weights for PSE and SE and deploying two 
distinct models leads to unnecessary memory and computational overhead, which is inefficient for industrial applications. 
This highlights the urgent need for a unified model capable of addressing both tasks effectively within a single framework.

Although in recent PSE society, few speaker embedding/ encoder-free PSE techniques have been proposed \cite{sef-net,yangxue,parnamaa2024personalized}, 
they remain incapable of addressing conventional SE tasks within their PSE frameworks due to the 
specialized operations for speaker enrollment and the interaction of mixture noisy speech representations.
In 2024, W. Zhang et al. organized the first URGENT (Universality, Robustness, and Generalizability for EnhancemeNT) 
Challenge \cite{zhang2024urgent} in the NeurIPS 2024 Competition Track, which aimed to 
develop universal speech enhancement models capable of unifying speech processing across diverse conditions.
This challenge emphasized the development of methods that could adaptively handle input speech 
with various distortions (corresponding to different SE subtasks) and input formats (e.g., sampling frequencies) 
in a range of acoustic environments, including those affected by noise and reverberation. 
Motivated by the URGENT Challenge, some new works have been proposed recently, 
such as AnyEnhance \cite{anyenhance}, which handles both speech and singing voices and supports denoising, 
dereverberation, declipping, super-resolution, and target speaker extraction within a unified framework.
Additionally, audio foundation models like UniAudio \cite{uniaudio} and Metis \cite{metis} have been developed to 
support several audio generation tasks covering different SE distortions, PSE tasks, text-to-speech, Lip2Speech, 
and more. Although these approaches can handle multiple speech enhancement tasks within a single model, 
many still require task-specific fine-tuning or are unable to manage multiple distortions simultaneously, 
and they often rely on large models that are challenging to deploy in low-resource industrial scenarios. 
Therefore, developing a single unified and lightweight framework that efficiently handles both SE and PSE tasks 
remains a fundamental and significant research objective.

Another important issue in PSE is the model robustness to enrollment speech variations, because 
the target speaker identity information in the enrollment speech is very critical for guiding the 
model to perform the high-quality target speech extraction. 
However, variations in enrollment speech are very complex, such as differences in speaking rate, 
emotional tone, semantic content or even background acoustic environments, can significantly affect 
PSE performance. For example, if the enrollment speech shares characteristics such as 
emotional tone, speaking rate, or intonation with the interfering speech in the mixture, 
the model may incorrectly extract the interfering 
speaker instead of the target speaker. As noted by \cite{target-confusion}, speaker confusion can 
occur when interfering speech closely resembles the target speaker, leading to ambiguous embeddings that 
misguide the network. This issue is amplified by extraneous factors in the enrollment speech, 
emphasizing the challenge of disentangling the speaker identity from irrelevant content. 
Addressing this is key to enhancing the robustness of target speech extraction systems.

Many speaker identity disentanglement approches have been proposed in the literature, 
for example, authors in \cite{self-supervised} have attempted to disentangle speaker identity from irrelevant attributes 
in the enrollment speech, but it relies on complex architectures with multiple encoder branches, 
reconstruction modules, and numerous loss functions for staged training. 
Such complexity can hinder the practical deployment of PSE systems, especially in real-time 
applications where computational resources are limited.
\cite{pbsrnn,x-sepformer} used conventional methods that rely on pre-trained speaker verification models 
for embedding extraction. Although the embeddings generated by these models achieve high accuracy in 
speaker verification tasks, they may suffer from domain mismatches, leading to suboptimal embeddings for PSE tasks. 
Similarly, the Transformer-based embedder proposed in \cite{centroid-estimation} enhances 
robustness by estimating speaker centroids. This approach refines the extracted embeddings by passing 
them through the Transformer-based embedder to generate new embeddings, which are then optimized 
using a loss function to align with the centroids of all enrollment speech. 
This process effectively isolates the speaker identity from irrelevant acoustic variations, 
such as noise or diverse acoustic conditions. However, the method's reliance on precise 
centroid estimation makes it sensitive to the quality of enrollment speech. 
Performance may degrade when enrollment speech is noisy, too short, or contains highly
variable acoustic characteristics, highlighting the need for more robust speech disentanglement mechanisms.

To address these challenges, we propose two novel models: \textbf{USEF-PNet} (\textbf{U}nified Architecture on SEF-PNet) 
and \textbf{DSEF-PNet} (Unsupervised Speech \textbf{D}isentanglement on SEF-PNet). Both models build upon our previously 
proposed speaker embedding/ encoder-free PSE system, SEF-PNet \cite{sefpnet}. 
\textbf{USEF-PNet} unifies PSE and SE tasks into a single model, by introducing a simple yet effective 
model training strategy. This enables the model to simultaneously handle both tasks, 
achieving performance that is either better or equal to that of task-specific models, 
while offering practical advantages in memory and computational efficiency for industrial deployment.
\textbf{DSEF-PNet} improves the robustness of the PSE model under varying enrollment speech conditions by using a novel 
unsupervised enrollment speech disentanglement mechanism. This mechanism separates speaker identity 
information from other irrelevant factors in the enrollment speech, enabling the model to focus on the 
target speaker's characteristics and improving its overall robustness.
Building upon \textbf{DSEF-PNet}, 
we also explore the Long-Short Enrollment Pairing (\textbf{LSEP}) 
strategy, which aims to leverage the advantages of long enrollment speech to mitigate the limitations of short enrollment speech and is expected to benefit short enrollment scenarios.
Our three key innovations of this study are summarized as follows: 

\begin{enumerate}[itemsep=10pt]

\item \textbf{Unified Architecture:} 
The proposed \textbf{USEF-PNet} integrates PSE and SE tasks within a single framework through a 
straightforward yet efficient training strategy. In this approach, each training batch is especially 
organized to include both SE and PSE input conditions. Specifically, for the SE task,
the enrollment speech is replaced with zeros input to render it ineffective, 
while for the PSE task, the valid enrollment speech is utilized as usual. This strategy enables 
the model to simultaneously handle both tasks, achieving performance comparable to or even 
surpassing that of task-specific models on the Libri2Mix and VoiceBank-Demand datasets, 
while also offering practical efficiency advantages for industrial applications.

\item \textbf{Unsupervised Speech Disentanglement:} 
The proposed \textbf{DSEF-PNet} enhances the robustness of PSE models to substantial variations in 
enrollment speech, such as changes in emotion, speaking rate, semantic content, and noise, 
through a novel unsupervised speech disentanglement approach. 
During \textbf{DSEF-PNet} training, two distinct enrollment speech samples are paired for each input mixture, 
and a consistency constraint is imposed to ensure that their outputs are the same. 
This mechanism implicitly disentangles speaker identity information from irrelevant content, 
thereby isolating the essential target speaker characteristics. 
The enrollment speech pairing operation, reinforced by these consistency constraints, 
enables the model to adapt effectively to a wide range of enrollment conditions, 
resulting in significant and consistent performance gains across multiple PSE scenarios in the Libri2Mix dataset.

\item \textbf{Long-Short Enrollment Pairing (LSEP):} In PSE tasks, performance often degrades significantly when the enrollment speech is extremely short. To investigate potential solutions to this issue, we explore a Long-Short Enrollment Pairing (\textbf{LSEP}) strategy within DSEF-PNet. \textbf{LSEP} pairs an extremely short enrollment sample with a long one as distinct enrollment clues for the same input noisy mixture during training, aiming to provide richer speaker guidance. However, experimental results show that this strategy does not lead to noticeable improvement. Random selection of enrollment duration yields better overall performance across different scenarios.

\end{enumerate}

\vspace{10pt}
The remainder of this paper is organized as follows. Section \ref{sec:review} reviews the 
related works, Section \ref{ssec:SEF-PNet} revisits our previously proposed SEF-PNet, 
The proposed methods are presented in Section \ref{sec:proposed}, including the USEF-PNet, DSEF-PNet, and the exploration of LSEP strategy. 
Section \ref{sec:exps} describes the experimental setups and  
Section \ref{sec:rst} presents the results and discussions, followed by conclusions.

\section{Review of Related Works}
\label{sec:review}

This study aims to build a well-established speech enhancement framework that can effectively 
handling both PSE and conventional SE tasks in real-world scenarios, related works 
including four main directions: the speaker encoder/embedding-free PSE techniques, 
the enrollment speech disentanglement methods, the universal speech enhancement architectures, 
and the approaches for dealing with short-duration enrollment speech.

\subsection{Speaker Encoder/Embedding-free PSE}

The core distinction between PSE and SE tasks is the incorporation of enrollment speech.
In PSE, the target speaker's identity, derived from enrollment speech, acts as a critical cue 
to guide the extraction of the target speaker’s voice from a multi-speaker noisy mixture, 
entailing two inputs and one output. In contrast, conventional SE operates without explicit 
target speaker guidance, its primary function is to perform denoising and dereverberation, 
enhancing speech perception and intelligibility, which involves one input and one output.
Traditional PSE approaches typically utilize target speaker embeddings, 
either extracted from pre-trained speaker verification models \cite{ECAPA-TDNN, resnet}, or generated via 
self-designed speaker encoders \cite{td-speakerbeam,spex,spex+,mc-spex,sdpccn}, 
as identity clues to isolate high-quality target speech from noisy mixtures. 
Although effective in many scenarios, these methods often suffer from increased model complexity, 
large parameter sizes, and slower inference speeds, which pose challenges for industrial deployment.

To address these issues, recent research has explored speaker encoder/embedding-free approaches, 
which bypass the dependency on explicit embeddings. For instance, the authors in \cite{sef-net} 
utilize weight-sharing encoders to jointly process enrollment and mixture speech, enabling direct 
extraction of target speaker information and eliminating the need for a speaker verification model. 
Similarly, the authors in \cite{yangxue} focus on interaction mechanisms in the time-frequency domain, 
directly modeling relationships between enrollment and mixture speech, thereby avoiding both speaker 
verification models and self-designed speaker encoders. Additionally, the authors in 
\cite{parnamaa2024personalized} propose extracting speaker embeddings internally within the speech 
enhancement model itself, further removing reliance on pre-trained speaker verification models. 
While these methods demonstrate promising potential, they still face challenges such as 
scalability and limited utilization of enrollment speech information.

Building on these advancements, our prior work \cite{sefpnet}, introduced SEF-PNet, a speaker encoder/ embedding free 
framework designed to simplify the sDPCCN \cite{sdpccn} while delivering competitive performance.
SEF-PNet achieves this through the proposed context interaction and iterative feature integration methods, 
which provide effective, implicit guidance for target speech extraction and enhancement in PSE,
eliminating the need for pre-trained speaker verification models or self-designed speaker encoders.
The competitive results obtained on the Libri2Mix dataset across three PSE conditions, as demonstrated in 
\cite{sefpnet}, underscore the framework's effectiveness and its potential for real-world applications.
Therefore, we adopt SEF-PNet as the backbone architecture in this study.

\subsection{Enrollment Speech Disentanglement}

The extraction of target speaker identity information from enrollment speech is critical for guiding PSE models, 
yet its effective utilization remains challenging due to issues such as speaker confusion and embedding robustness. 
As highlighted in \cite{target-confusion}, global attributes present in the enrollment speech, 
such as emotion, speaking rate, or prosody, can mislead 
the extraction network, especially in scenarios where interfering speakers possess  
similar vocal characteristics. This challenge has led to significant research efforts 
aimed at enhancing the robustness of PSE systems with respect to variations in enrollment speech.

Most conventional PSE systems predominantly rely on embeddings derived from pre-trained 
speaker verification (SV) models \cite{ECAPA-TDNN,resnet}. Many speech disentanglement methods 
have been proposed to augment the SV model training and produce robust 
speaker embeddings. For instance, \cite{tu2024contrastive} introduced a sequential disentangling 
contrastive learning framework that combines a disentangling sequential variational autoencoder 
with traditional contrastive learning to remove the influence of language content.  
In \cite{sunzhang2023}, a noise-disentanglement metric learning approach was proposed to suppress the 
speaker-irrelevant noisy components and build a noise-invariant embedding space, thereby producing 
robust speaker embeddings in noisy environments. Additionally, \cite{yilu22} presented an 
InfoMax–DSAN framework to disentangle domain-specific features from domain-invariant speaker features, 
improving the domain robustness of SV systems, While in \cite{Tianchi23}, authors 
proposed RecXi,  a self-supervised disentangling framework aimed at separating speaker and content 
representations in SV. Although these approaches effectively address the robustness of speaker embedding 
extraction within the SV domain, they often yield embedding patterns that do not align with the specific 
requirements of the PSE task, resulting in suboptimal performance.

An alternative line of research in PSE systems employs self-designed speaker-encoders to generate 
target speaker embeddings for model guidance. However, relatively few studies have focused on 
disentangling speaker identity information directly from enrollment speech. For example,  
in \cite{self-supervised}, the authors removed local semantic information deemed irrelevant
to speaker identity by proposing a multi-branch encoder-based enrollment speech reconstruction 
disentangling module, utilizing complex loss functions in a two-stage model training process. 
In \cite{zhaogao20}, speaker-encoders were trained using a multi-class cross-entropy loss 
with speaker identity labels and jointly optimized with the target speech extraction network.
Although these speaker-encoder dependent approaches have effectively reduced interference 
from unrelated attributes, their complexity and computational demands may not be 
suitable for real-world industrial scenarios. 

In contrast to both pre-trained SV model-based and self-designed speaker encoder-based PSE systems, 
to the best of our knowledge, no prior research has addressed enrollment speech disentanglement 
in speaker encoder/embedding-free PSE or target speaker extraction (TSE) systems. 
Given the rapid advancements in speaker encoder/embedding-free PSE techniques, it is both necessary 
and meaningful to explore enrollment speech disentanglement strategies that enhance enrollment speech 
robustness without incurring additional computational overhead.

\subsection{Universal Speech Enhancement}
\label{subsec:use}

Conventional speech enhancement models have traditionally  been trained and evaluated under 
specific conditions, such as fixed datasets, sampling rates, and well-controlled training environments. 
These models are typically designed for a single task, such as noise suppression, dereverberation, 
or target speaker extraction (TSE). As a result, they lack the ability to address multiple enhancement tasks 
simultaneously or to generalize effectively to more complex and diverse acoustic environments. 
Therefore, in recent years, some new works start to explore unified SE frameworks that  
improve a model's universality in handling real-world SE challenges.
For example, in \cite{Serra22, yenjulu22}, diffusion-based generative model was proposed to address different 
distortions (eg. reverb, clipping, codec artifacts, etc) with a single model, while  
VoiceFixer \cite{Haohe22} has been introduced to restore speech degraded by  
multiple distortions (e.g., noise, reverberation, and clipping) and
upscale low bandwidth noisy speech to 44.1 kHz full-bandwidth high-fidelity speech. 
Moreover, recent efforts in \cite{zhangwsaijo23,zhangqian24} have demonstrated  the 
possibility of building a single system to 
handle various input formats, including different sampling frequencies and numbers of microphones.

In 2024, the URGENT Challenge \cite{zhang2024urgent} was launched to establish a comprehensive benchmark 
that spans a wide range of SE conditions and facilitates systematic comparisons between 
state-of-the-art discriminative and generative SE methods regarding their generalizability. 
Driven by the advancements showcased in the URGENT Challenge, 
several new speech foundation models, such as AnyEnhance \cite{anyenhance}, UniAudio\cite{uniaudio}, and Metis\cite{metis}, 
have recently emerged. These models are designed not only to deliver versatile speech enhancement 
capable of robust performance across diverse real-world applications but also to extend their 
capabilities to cover a broad array of speech generation tasks, 
including TTS, Lip2Speech, TSE, singing voice enhancement, etc. Although these models offer the potential to 
handle multiple speech enhancement tasks within a single framework, many still require task-specific fine-tuning 
or exhibit difficulties in simultaneously handling multiple distortions.
Additionally, their reliance on large-scale models poses challenges for deployment in 
resource-constrained industrial environments. 
Therefore, developing a unified and lightweight framework that can efficiently address both SE and PSE tasks 
remains a key and crucial research challenge.

\subsection{Short-Duration Enrollment}

Previous studies have consistently demonstrated that longer enrollment utterances tend to 
yield improved performance in target speech extraction \cite{spex++, spex+}. 
However, practical applications often impose constraints on enrollment duration; 
for example, mobile devices typically restrict enrollment to brief utterances, 
such as wake-up words, to maintain user engagement. In the field of speaker verification (SV), 
various methods have been proposed to address the challenges associated with short-duration enrollment. 
Such as, in \cite{Zeinali2021}, the Short-duration Speaker 
Verification (SdSV) Challenge was organized to focus the analysis and exploration of 
new ideas for short-duration speaker verification. 
Authors in \cite{wangshuai19} proposed a cascade embedding learning strategy with 
deep discriminant analysis to enhance SV under short-duration conditions. 
A ERes2NetV2 computational efficiency framework was designed in \cite{cheny2024} to 
effectively capture features from short-duration utterances, while in \cite{huangz25}, the speech synthesis was 
applied to augment enrollment speech, thus enhancing the short-duration SV system performance.

In contrast to the SV domain, the challenge of short-duration enrollment has received comparatively 
little attention in PSE or TSE tasks.  Recent advances in this area include methods such as 
Spex++ \cite{spex++} and the Voice Extractor-Voice Extractor (VE-VE) framework \cite{veven}. 
Spex++ improves target speaker extraction by progressively refining the speech extraction process, 
which enhances performance even with ultra-short enrollment speech. However, the method heavily relies 
on the accuracy of the initial extraction stage, which can limit its overall performance. 
Meanwhile, the VE-VE framework utilizes an RNN to synchronously update the enrollment speech representation, 
effectively addressing the challenges posed by ultra-short enrollment.
However, the reliance on RNNs limits the approach's scalability and applicability in other scenarios.

\vspace{0.3cm}

In summary, existing research on Personalized Speech Enhancement (PSE) has made significant strides 
in addressing speaker encoder/embedding-free approaches, enrollment speech disentanglement, 
universal speech enhancement, and short-duration enrollment challenges. However, several issues remain 
unresolved: 1) while speaker encoder/embedding-free methods have shown promising results, 
further advancements are needed to develop more effective and scalable solutions; 
2) Simpler and more universal strategies for disentangling enrollment speech across various models 
have not been fully explored; 3) while universal SE or speech foundation models have made progress, 
they have yet to incorporate PSE tasks into a efficient and light-weighted unified framework, especially in 
resource-constrained industrial environments. Finally, there is a lack of sufficient research into 
enhancing an unified SE and PSE system performance under extreme short-duration enrollment speech conditions. 
In this work, we focus on addressing these issues through two key innovations:
USEF-PNet, which tackles the universal PSE-SE problem; and DSEF-PNet, which introduces a simple and effective unsupervised method for enrollment speech disentanglement.
In addition, we explore the LSEP strategy as a potential means to improve performance under extreme short-duration enrollment conditions.

\begin{figure*}[ht]
\centering
\setlength{\abovecaptionskip}{0cm}
\includegraphics[width=0.92\textwidth]{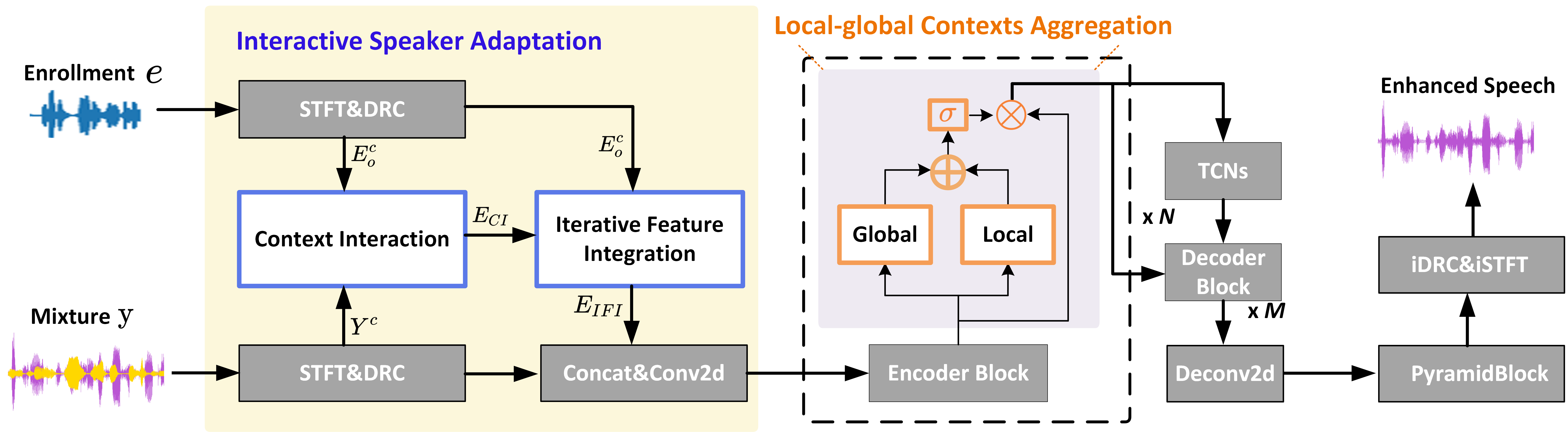}
\caption{Overview of our previously proposed SEF-PNet model. All colored blocks highlight the 
key contributions over the original sDPCCN.}
\label{fig:isalca}
\end{figure*}

\section{Revisit SEF-PNet}
\label{ssec:SEF-PNet}

In this section, we revisit the SEF-PNet architecture, which builds upon the sDPCCN \cite{sdpccn} framework by 
eliminating the speaker encoder module and redesigning the front-end to create a speaker encoder/embedding-free 
system. The key innovation of SEF-PNet lies in its integration of the Interactive Speaker Adaptation (ISA) 
module, which combines Context Interaction (CI) and Iterative Feature Integration (IFI) to operate on both mixture 
and enrollment speech. This approach effectively removes the need for an explicit speaker encoder while maintaining 
strong performance in PSE. Additionally, the Local-global Contexts Aggregation (LCA) module is introduced to enhance 
the robustness of the system by incorporating both local and global contextual information. These advancements 
significantly improve the model's PSE performance while maintaining a reduced model complexity. 
In the following, we revisit our SEF-PNet from three key aspects: the overall architecture, ISA module and the 
LCA module.

\textbf{Model Architecture:} The structure of SEF-PNet, as illustrated in Fig.\ref{fig:isalca}, builds upon our previous work, sDPCCN, which has already demonstrated competitive performance in end-to-end TSE tasks. Using sDPCCN as the backbone, we implemented several key modifications to achieve a speaker encoder-free design. First, we removed the original speaker encoder module, simplifying the architecture and focusing on the direct utilization of enrollment speech, as shown in the ISA module (left part of Fig.\ref{fig:isalca}). Second, we integrated the LCA module into the original Encoder Block to exploit both local and global contextual information in the high-level acoustic representations. 
In addition, we reduced the number of Encoder Blocks from 7 to 6 and shortened the frame window length from 64ms to 32ms. These changes preserved the model's performance while reducing its size by approximately 0.7M parameters. The others remain consistent with those in the original sDPCCN. Finally, as shown in Fig.\ref{fig:isalca}, in SEF-PNet, the entire encoder now consists of an ISA adaptation module followed by 6 Encoder Blocks, each enhanced with the LCA module, significantly boosting PSE performance while maintaining much lower model complexity.

\textbf{Interactive Speaker Adaptation (ISA):} The CI module aligns the mixture speech with the enrollment speech in the time-frequency (T-F) domain. However, this process may distort the enrollment speech and cause loss of original speaker-specific information. To address this issue, the IFI module in ISA uses both the transformed enrollment speech and the output of CI. Through two iterative steps, it progressively integrates these features, gradually recovering the target speaker's characteristics while minimizing distortion from the noisy speech. By iterating this process, the model is able to effectively enhance the target speaker clues without relying on external embeddings. 

\textbf{Local-global Contexts Aggregation (LCA):}
The LCA module utilizes two parallel attention mechanisms: Global Attention (GA) and Local 
Attention (LA), to capture global, speaker-independent features as well as localized, 
temporal aspects essential for speech enhancement. The fusion of global and local 
context information via these attention mechanisms leads to better feature recalibration, 
improving the model's overall performance in PSE tasks.

\vspace{0.2cm}
As demonstrated in our previous experiments, extensive results show that SEF-PNet achieves state-of-the-art 
performance in PSE across different conditions. For further details on the implementation and
mathematical formulation, the reader is referred to our earlier work in \cite{sefpnet}.

\begin{figure}[!htbp]
\centering
\setlength{\abovecaptionskip}{0cm}
\includegraphics[width=0.45\textwidth]{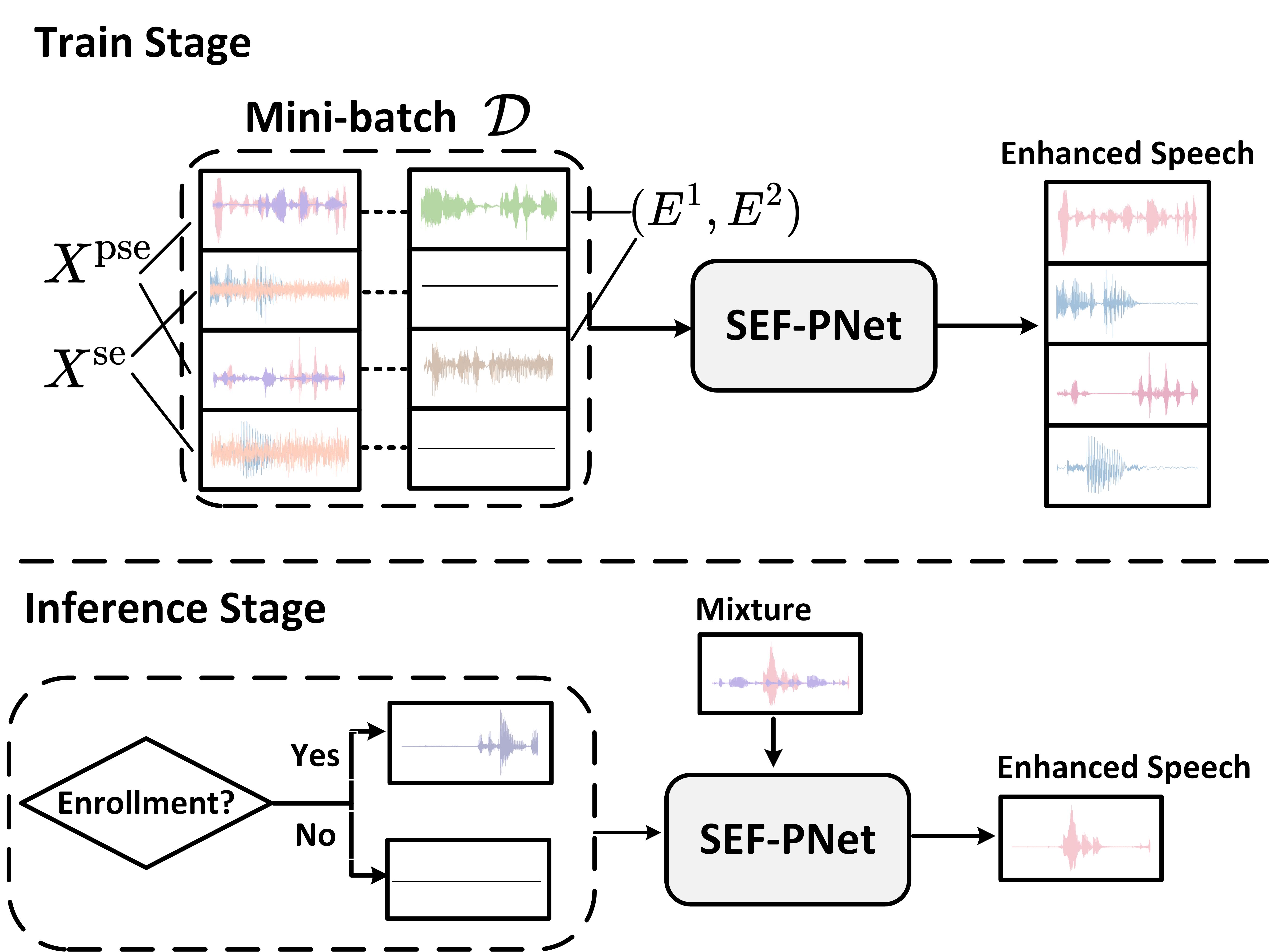}
\caption{Structure of Unified Architecture on SEF-PNet (USEF-PNet).}
\label{fig:usef-pnet}
\end{figure}

\begin{figure*}[t]
\centering
\setlength{\abovecaptionskip}{0cm}
\includegraphics[width=1.0\textwidth]{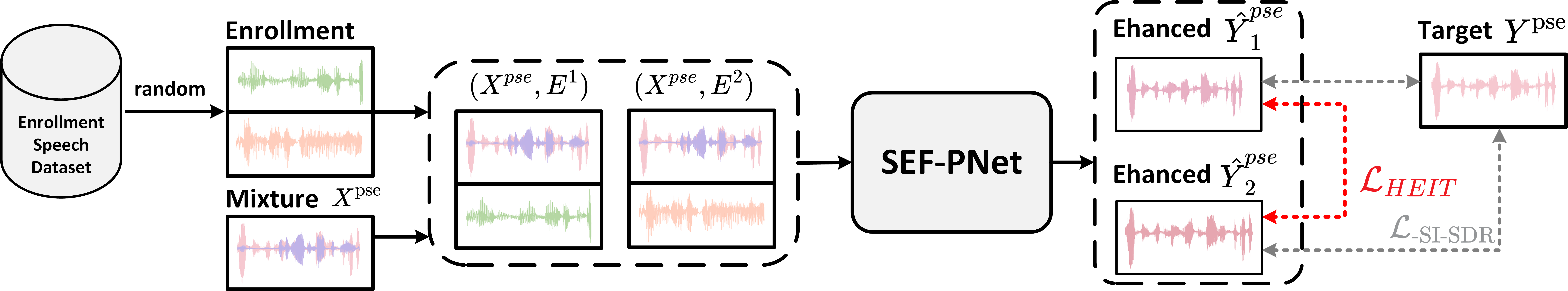}
\caption{\centering The whole training framework of DSEF-PNet with heterogeneous enrollment invariant training (HEIT).}
\label{fig:dsef-pnet}
\end{figure*}

\section{Proposed Methods}
\label{sec:proposed}

In this study, we propose two novel methods to enhance the universality and robustness of our previously proposed SEF-PNet for speaker-embedding/ encoder-free personalized speech enhancement.
The first method, USEF-PNet, is introduced to integrate both conventional SE and PSE into a single model, 
without altering the original architecture of SEF-PNet. 
The second method, DSEF-PNet, incorporates an unsupervised enrollment speech disentanglement mechanism, which improves the model's robustness to various acoustic variations in the enrollment speech. 
Additionally, we explore the integration of the LSEP approach within DSEF-PNet as a potential strategy for addressing performance challenges under short-duration enrollment conditions.
Lastly, we explore and evaluate the combination of USEF-PNet and DSEF-PNet on both SE and PSE tasks. Detailed implementations of USEF-PNet, DSEF-PNet, and LSEP are provided in the following subsections.

\subsection{USEF-PNet}
\label{sub:usefnet}

Most existing speech enhancement models are designed for specific tasks, such as noise suppression or 
target speaker extraction (TSE), and often face challenges in generalizing across diverse scenarios. 
Recent advancements in universal models or speech foundation models have shown potential; 
However, they are not yet suitable for industrial applications due to their high model complexity 
or the need for task-specific fine-tuning, etc. To address this issue, we introduce the USEF-PNet, 
a model that seamlessly integrates both personalized speech enhancement and conventional SE 
within a single framework, providing an efficient and versatile solution for universal speech enhancement.

The training and inference procedures of USEF-PNet are illustrated in Fig.\ref{fig:usef-pnet}. It is clear to 
see that our proposed USEF-PNet has the same model architecture with our previously proposed SEF-PNet. 
The only difference between the two models lies in their training strategies. 
While SEF-PNet is trained specifically for PSE tasks, using paired noisy mixture speech 
and target speaker enrollment speech as inputs, USEF-PNet must accommodate both PSE and SE 
task compatibility. This is because conventional SE tasks do not provide enrollment speech.
Specifically, as shown in the upper part of Fig.\ref{fig:usef-pnet}, in the training stage, 
each training mini-batch $\mathcal{D}$ includes samples from both SE and PSE training sets, 
organized as follows: 

\begin{equation}
\begin{aligned}
&\mathcal{D} = \{\mathcal{D}_{pse} \cup  \mathcal{D}_{se}\} \\
&\mathcal{D}_{pse} =  \bigcup_{i=1}^{M} \left\{(X_{i}^{\mathrm{pse}}, E_{i}), Y_{i}^{\mathrm{pse}}\right\}  \\ 
&\mathcal{D}_{se} = \bigcup_{j=1}^{N} \left\{(X_{j}^{\mathrm{se}}, \mathbf{0} ),  Y_{j}^{\mathrm{se}} \right\}
\end{aligned}
\label{eq:eq1}
\end{equation}
where $\mathcal{D}_{pse}$ and $\mathcal{D}_{se}$ represent the $M$ and $N$ randomly selected 
samples from the PSE and SE training datasets, respectively. Here, 
$X_{i}^{\mathrm{pse}}$ and $X_{j}^{\mathrm{se}}$ correspond to the $i$-th and $j$-th input mixture 
and noisy training samples from PSE and SE training sets, respectively. The corresponding clean 
speech targets  (ground-truth) are denoted as $Y_{i}^{\mathrm{pse}}$ and $Y_{j}^{\mathrm{se}}$. 
For the PSE training samples, $E_{i}$ represents the enrollment speech for 
$X_{i}^{\mathrm{pse}}$, while for SE training samples, the enrollment speech is absent, denoted as $\mathbf{0}$.

With each $\mathcal{D}$, the USEF-PNet model is trained by minimizing the traditional negative scale-invariant 
signal-to-distortion ratio (SISDR) loss function \cite{sisdr}, as defined by:

\begin{equation}
\mathcal{L}_{USEF-PNet} = \mathcal{L}_{\text{-SISDR}} (\hat{Y}, Y)
\label{eq:eq2}
\end{equation}
where $ \hat{Y} $  represents the predicted enhanced speech output from USEF-PNet, with
$\hat{Y} \in \{ \hat{Y}^{pse}, \hat{Y}^{se}\}$ being the enhanced speech for either the PSE or SE task. 
Similarly, $Y\in \{Y^{pse}, Y^{se}\}$  is the corresponding target clean speech.
By incorporating both PSE and SE tasks within a unified framework, USEF-PNet achieves a 
highly efficient and flexible solution for universal speech enhancement.

During inference stage, as shown in the below part of Fig.\ref{fig:usef-pnet}, 
the system first checks whether valid enrollment speech is available. If it is, the model takes 
both the noisy mixture speech and the enrollment speech as inputs to generate the enhanced speech 
$\hat{Y}^{pse}$. If no enrollment speech is provided, the system defaults to using a zero-valued dummy 
enrollment speech $\mathbf{0}$, and processes the noisy speech alone, and produce the enhanced output 
$\hat{Y}^{se}$. This flexibility ensures that USEF-PNet can handle both personalized speech 
enhancement and conventional speech enhancement tasks, adapting to real-world scenarios where 
enrollment speech may or may not be available.
It is worth noting that the principle behinds simply setting the enrollment speech to $\mathbf{0}$
in the SE scenario within USEF-PNet lies in the fact that we employ a concatenation operation between
the enrollment speech representation $E_{IFI}$ and the STFT\&DRC transformed input mixture features,
as illustrated in Fig.\ref{fig:isalca}.

\subsection{Unsupervised Enrollment Speech Disentanglement (DSEF-PNet)}
\label{subsec:uesd}

A stable and reliable target speaker identity representation is crucial for 
guiding PSE systems in extracting high-quality target speech from noisy mixtures. 
However, this identity information often suffers from the issue of target confusion \cite{target-confusion}, 
because of the irrelevant acoustic, prosodic, or semantic content included the enrollment speech, thus  
misleading the speech extraction network and resulting in inaccurate outcomes. 
Therefore, disentangling the enrollment speech to extract robust speaker identity 
information is essential for effective PSE. While previous works have primarily focused on PSE 
systems that rely on target speaker embeddings extracted from pre-trained speaker verification 
models or self-designed speaker encoders, there has been a lack of methods addressing 
enrollment speech disentanglement in speaker-embedding/ encoder-free PSE systems.

In this work, we propose a novel enrollment speech disentanglement approach, 
termed Heterogeneous Enrollment Invariant Training (HEIT), to improve the robustness 
of SEF-PNet in an unsupervised manner. We refer to the resulting system as DSEF-PNet 
for simplicity. The overall training framework of DSEF-PNet is demonstrated in 
Fig.\ref{fig:dsef-pnet}. Specifically, unlike the conventional PSE methods that use 
one-to-one noisy mixture and enrollment speech inputs at a time, 
the mini-batch of training data for DSEF-PNet is oragnized as follows:

\begin{equation}
\mathcal{D}_{DSEF-PNet} =  \bigcup_{i=1}^{K} \left\{(X_{i}^{\mathrm{pse}}, E_{i}^{1}, E_{i}^{2}), \ \ Y_{i}^{\mathrm{pse}}\right\}  
\label{eq:eq3}
\end{equation}
where $(X_{i}^{\mathrm{pse}}, E_{i}^{1}, E_{i}^{2})$ are the inputs to the SEF-PNet model, and $Y_{i}^{\mathrm{pse}}$ is 
the corresponding ground-truth. The two distinct enrollment speech recordings $ E_{i}^{1}$ and 
$ E_{i}^{2} $ are with the same target speaker's identity but include
heterogeneous identity-irrelevant contents.

As shown in Fig.\ref{fig:dsef-pnet}, for each input  $X^{\mathrm{pse}}$, the paired inputs 
 $(X^{\mathrm{pse}}, E^{1})$ and $(X^{\mathrm{pse}}, E^{2})$ 
are processed in parallel by the SEF-PNet model, yielding two enhanced speech outputs 
$\hat{Y}_{1}^{\mathrm{pse}}$ and $\hat{Y}_{2}^{\mathrm{pse}}$, 
respectively. By aligning these two enhanced outputs, 
$\hat{Y}_{1}^{\mathrm{pse}}$ and $\hat{Y}_{2}^{\mathrm{pse}}$.
we implicitly equip SEF-PNet with the ability to disentangle the target speaker identity. 
This force-alignment is achieved using the final output-level HEIT loss defined as, 
\begin{equation}
\mathcal{L}_{HEIT}=  \|\hat{Y}_{1}^{\mathrm{pse}} - \hat{Y}_{2}^{\mathrm{pse}} \|_1
\label{eq:eq4}
\end{equation}
Where $\hat{Y}_{*}^{\mathrm{pse}} $ denotes the complex spectrogram of the enhanced speech output, 
and the L1-norm measures the mean absolute error (MAE) between the two outputs.

Thus, the total loss for 
training the proposed DSEF-PNet model is then formulated as, 
\begin{equation}
\begin{split}
\mathcal{L}_{DSEF-PNet}  & =  \mathcal{L}_{\text{-SISDR}} (\hat{Y}_{1}^{\mathrm{pse}}, Y^{\mathrm{pse}}) \\
& + \mathcal{L}_{\text{-SISDR}} (\hat{Y}_{2}^{\mathrm{pse}}, Y^{\mathrm{pse}}) 
 + \lambda  \cdot \mathcal{L}_{HEIT}
\end{split}
\label{eq:eq5}
\end{equation}
where $\lambda$ is a tunable weight for the HEIT loss. 
As shown in in Eq.(\ref{eq:eq5}), together with the $\mathcal{L}_{HEIT}$, 
the implicitly enrollment speech disentanglement is further enhanced by 
the force-aligned $(\hat{Y}_{1}^{\mathrm{pse}}, Y^{\mathrm{pse}})$  and 
$(\hat{Y}_{2}^{\mathrm{pse}}, Y^{\mathrm{pse}})$ 
using the standard negative SISDR loss functions, thus completing the whole 
heterogeneous enrollment invariant training.

This combined loss, $\mathcal{L}_{DSEF-PNet}$, ensures consistent enhancement of target speech regardless 
of variations in enrollment recordings, enabling the model to focus on speaker-identity-related 
features while remaining invariant to irrelevant factors such as emotion, tone, prosody, or even noise. 
Importantly, DSEF-PNet does not introduce additional parameters or inference-time overhead beyond that of 
standard PSE in SEF-PNet, making it highly practical for real-world applications. 
By enhancing robustness to enrollment speech variability, DSEF-PNet significantly improves the generalization 
capability of the overall speech enhancement process.

\subsection{Long-Short Enrollment Pairing (LSEP)}
\label{subsec:lsep}

Typically, in both SV and PSE systems, the enrollment speech is expected to be of sufficient 
length to accurately capture the speaker’s characteristics. However, in real-world applications, 
users often provide short-duration enrollment speech (e.g., 1-2 seconds), which can lead to 
suboptimal model performance. This is particularly problematic for systems that are trained only on 
long-duration enrollment speech, as they fail to generalize well to short-duration scenarios. 
Therefore, it is crucial to develop strategies that enhance model robustness, enabling it to 
handle both short and long enrollment speech effectively.

In this study, build upon the DSEF-PNet, we explore a Long-Short Enrollment Speech Pairing 
(LSEP) strategy to further improve its enrollment speech duration robustness. 
The LSEP is implemented in a very simple manner, just by replacing the two parallel 
heterogeneous enrollment speech recordings in Eq.(\ref{eq:eq3}) with one 
short-duration $E_{i}^{short}$ (eg., 2 seconds) and one long-duration enrollment $E_{i}^{long}$ (eg., 10 seconds) as, 
\begin{equation}
(E_{i}^{1}, E_{i}^{2}) = (E_{i}^{short}, E_{i}^{long})
\label{eq:eq6}
\end{equation}

The loss function of DSEF-PNet is then updated as follows:
\begin{equation}
\begin{split}
\mathcal{L}_{DSEF-PNet} & =   \mathcal{L}_{\text{-SISDR}} (\hat{Y}_{short}^{\mathrm{pse}}, Y^{\mathrm{pse}}) \\ 
& + \mathcal{L}_{\text{-SISDR}} (\hat{Y}_{long}^{\mathrm{pse}}, Y^{\mathrm{pse}})  + \lambda  \cdot \mathcal{L}_{HEIT}
\end{split}
\label{eq:eq7}
\end{equation}
where $\hat{Y}_{short}^{\mathrm{pse}}$ and $\hat{Y}_{long}^{\mathrm{pse}}$ 
are the enhanced speech outputs obtained using
$E^{short}$ and $ E^{long}$, respectively. 

The principle behind LSEP is to force the DSEF-PNet model to learn a consistent speaker identity 
representation regardless of enrollment duration. By aligning the outputs generated from a 
short and a long enrollment sample through the HEIT loss, the model is encouraged to focus on 
identity-relevant features and become invariant to the differences in duration and extraneous content. 
This, in turn, is intended to improve the model’s generalization capability, ensuring robust performance even 
when only short enrollment speech is available.

In summary, integrating LSEP within DSEF-PNet improves the model's robustness across varying enrollment durations compared to SEF-PNet. However, the performance is not as strong as when two random enrollment durations are used. Nonetheless, LSEP remains a valuable alternative for enhancing performance when compared to training exclusively with short enrollment speech. This approach aligns with the underlying principles of HEIT, focusing on consistent identity information.

\section{Experimental Setup}
\label{sec:exps}

\subsection{Datasets}

All our experiments are performed on two publicly available datasets, the Libri2Mix \cite{libri2mix} and 
VoiceBank-DEMAND \cite{voice-demand}. The Libri2Mix is used for evaluating the PSE systems, and the 
VoiceBank-DEMAND is used for conventional SE tasks. Libri2Mix dataset covers three public PSE conditions: 
1) 1-speaker+noise (\texttt{mix\_single}): the mixture signal with only a single target speaker and background noise;   
2) 2-speaker (\texttt{mix\_clean}): the mixture with a target speaker and one interfering speaker without any 
additional noise; 3) 2-speaker+noise (\texttt{mix\_both}): the mixture with a target speaker, one interfering speaker 
and background noise. The \texttt{mix\_*} names are original PSE conditions in Libri2Mix, For clarity, we use the terms 
`1-speaker+noise', `2-speaker' and `2-speaker+noise' as representative labels instead of the original ones. In each 
condition, the training set includes 13,900 utterances from 251 speakers, while both the development and test sets 
contain 3,000 utterances from 40 speakers each, with all mixtures simulated in the `minimum' mode. All mixtures are 
resampled to 8 kHz. Note that only the first speaker is taken as the target speaker during all training mixture data 
simulation, unless otherwise specified.

The original training set of VoiceBank-DEMAND consists of 11,572 utterances from 28 speakers across 
four signal-to-noise ratios (SNRs): 15 dB, 10 dB, 5 dB, and 0 dB. For validation, we randomly selected 
two speakers (p286 and p287) from the training set, resulting in a validation set of 905 utterances. 
The test set, consisting of 824 utterances, includes two speakers and four different SNRs: 17.5 dB, 
12.5 dB, 7.5 dB, and 2.5 dB. All data, including training, validation, and testing, is resampled to 8 kHz.

\begin{table*}[t] 
\renewcommand\arraystretch{1.2}
\caption{Unified Architecture Results on Libri2Mix (2-speaker,1-speaker+noise condition), 
with Training Data Size / 2. The `2-speaker' and `1-speaker+noise' condition are used to represent  
the PSE and conventional SE tasks, respectively. }
\label{tab:Unvi-libri2mix}
\centering
\scalebox{1.0}{
    \begin{tabular}{l|c|c|ccc|ccc}
    \toprule
    \multirow{3}{*}{\textbf{Training Data}} 
    & \multirow{3}{*}{\textbf{Methods}} 
    & \multirow{3}{*}{\textbf{Enrollment}} 
    & \multicolumn{6}{c}{\textbf{Evaluation sets}} \\
    & & & \multicolumn{3}{c}{\textbf{2-speaker}}
    & \multicolumn{3}{c}{\textbf{1-speaker+noise}} \\
    \cmidrule(lr){4-6} \cmidrule(lr){7-9}
    & & & \textbf{SISDR} & \textbf{PESQ} & \textbf{STOI} 
      & \textbf{SISDR} & \textbf{PESQ} & \textbf{STOI} \\
    \midrule
    \multirow{2}{*}{\textbf{(2-speaker)$_{1/2}$}} 
    & Mixture & - & -0.03 & 1.60 & 71.38 & - & - & - \\
    & SEF-PNet & Y & 10.90 & 2.64 & 87.04 & - & - & - \\
    \midrule
    \multirow{2}{*}{\textbf{(1-speaker+noise)$_{1/2}$}} 
    & Mixture & - & - & - & - & 3.32 & 1.75 & 79.51 \\
    & SEF-PNet & - & - & - & - & 13.80 & 2.94 & 91.61 \\
    \midrule
    \multirow{3}{*}{\textbf{(2-speaker)$_{1/2}$+(1-speaker+noise)$_{1/2}$}} 
    & \multirow{3}{*}{USEF-PNet} 
    & (Y, 0) 
    & \textbf{11.80} & \textbf{2.86} & \textbf{88.10} & \textbf{13.85} & \textbf{2.95}  & 91.67 \\
    & & (Y, 1) 
    & 11.67 & 2.86 & 87.93 & 13.63 & 2.89 & 91.20 \\
    & & (Y, random) 
    & 11.78 & 2.84 & 88.03 & 13.82 & 2.94 & \textbf{91.72} \\
    \bottomrule
    \end{tabular}}
\end{table*}

\subsection{Model Configurations}

The SEF-PNet architecture in our proposed USEF-PNet and DSEF-PNet consists of 7 Encoder Blocks, 
7 corresponding Decoder Blocks, a Temporal Convolutional Network (TCN) module, 
a PyramidBlock, and a Deconv2d layer. The Encoder Blocks are configured as follows: 
the first block contains a DenseBlock, the subsequent three blocks each incorporate DenseEncoders, 
and the final three blocks consist of only Conv2dBlocks. For the Decoder Blocks, the initial 
block features three Deconv2dBlocks, the next three blocks are composed of DenseDecoders, 
and the final block includes a DenseBlock. The TCN module is structured into two layers, 
each containing ten TCN blocks. The PyramidBlock integrates average pooling with four parallel 
branches, each employing a Conv2d layer followed by upsampling, with the outputs concatenated 
to produce the final result. For further details regarding the configurations of the Encoder 
Blocks and DenseEncoders, please refer to our previous work in \cite{sdpccn}.

In the Interactive Speaker Adaptation (ISA) and Local Context Attention (LCA) modules, 
all convolutional kernels and strides are set to 1, with channel upsampling or downsampling 
factors of $K = 1/32$ and $K = 4$, respectively. Specifically, the ISA module, which acts as 
the front-end, applies a dynamic range compression \cite{drc} factor of 0.5 to the STFT. 
Each EncoderBlock is followed by an LCA module. For further details on the ISA and LCA modules, see~\cite{sefpnet}.

The model is trained using the Adam \cite{adam} optimizer with an initial 
learning rate of 0.0005. The learning rate is adjusted by multiplying it by 
0.98 every two epochs for the first 100 epochs and by 0.9 in the last 20 epochs. 
Gradient clipping \cite{zhang2019gradient} is applied to limit the maximum L2-norm to 1. 
The training procedure lasts for up to 120 epochs.

\begin{table*}[t] %
\renewcommand\arraystretch{1.2}
\caption{Unified Architecture Results on the full libri2mix `2-speaker' condition + VoiceBank-Demand dataset.}
\label{tab:Unvi-libri2mixvctk}
\centering
\scalebox{1.0}{
    \begin{tabular}{l|c|c|ccc|ccc}
    \toprule
    \multirow{3}{*}{\textbf{Training Data}} 
    & \multirow{3}{*}{\textbf{Methods}} 
    & \multirow{3}{*}{\textbf{Enrollment}} 
    & \multicolumn{6}{c}{\textbf{Evaluation sets}} \\    
    & & & \multicolumn{3}{c}{\textbf{2-speaker}} 
    & \multicolumn{3}{c}{\textbf{VoiceBank-Demand}} \\
    \cmidrule(lr){4-6} \cmidrule(lr){7-9}
    & & & \textbf{SISDR} & \textbf{PESQ} & \textbf{STOI} 
      & \textbf{SISDR} & \textbf{PESQ} & \textbf{STOI} \\
    \midrule
    \multirow{3}{*}{\textbf{2-speaker}} 
    & Mixture & -
    & -0.03 & 1.60 & 71.38 & - & - & - \\
    & sDPCCN \cite{sdpccn} & Y 
    & 11.62 & 2.76 & 87.19 & - & - & - \\
    & SEF-PNet & Y 
    & 13.00 & \textbf{3.01} & 89.71 & - & - & - \\
    \midrule
    \multirow{3}{*}{\textbf{VoiceBank-Demand}}
    & Mixture & - 
    & - & - & - & 8.45 & 2.95 & 0.92 \\
    & MP-SENet \cite{mp-senet} & -
    & - & - & - & - & 3.50 & \textbf{0.96} \\
    & SEF-PNet & - 
    & - & - & - & 19.21 & 3.54 & 0.95  \\
    \midrule
    \textbf{2-speaker\ +\ VoiceBank-Demand}
    & USEF-PNet
    & (Y, 0) 
    & \textbf{13.00} & 2.99 & \textbf{89.74} & \textbf{19.69} & \textbf{3.59} & 0.94 \\
    \bottomrule
    \end{tabular}}
\end{table*}

\subsection{Evaluation Metrics}
We employ three evaluation metrics to assess the performance of our proposed methods on PSE and 
SE tasks, including the SISDR (dB)\cite{sisdr}, PESQ\cite{pesq} and STOI (\%) \cite{stoi}. 
The SISDR is used to measures the improvement in signal-to-distortion ratio, quantifying the 
separation quality of the target speech. The PESQ is used for evaluating the perceptual quality of 
the enhanced speech based on a psychoacoustic model, while the STOI is applied to 
assesses the speech intelligibility by computing the correlation between clean and enhanced 
speech spectrograms.

\section{Results and Discussions}
\label{sec:rst}
\subsection{Results for Unified Architecture}
\subsubsection{Results on Libri2Mix (2-speaker, 1speaker + noise) with training data size/2}

Table \ref{tab:Unvi-libri2mix} provides a unified evaluation of our proposed USEF-PNet 
and baseline systems under two different Libri2Mix conditions, the `2-speaker' (taken as the PSE task) and `1-speaker+noise' (taken as the SE task). Notably, only half of the original 
Libri2Mix data is used for these experiments. This choice was made to avoid speaker overlap 
between the PSE and SE conditions when we built USEF-PNet model: 
the speaker designated as the interference in the PSE condition may be taken as the target in the SE condition, and vice versa. By limiting the dataset, we ensure that the same speaker does not appear in conflicting roles (in the same training min-batch) across these two tasks.

In the `2-speaker' condition, enrollment speech is originally provided (denoted by “Y”), 
since extracting a specific target speaker from a mixture of two speakers inherently requires 
such information. In the `1-speaker+noise' condition, the system deals with a standard speech enhancement 
task, one speaker corrupted only by noise. However, in the proposed USEF-PNet system, 
we tried using (Y,0/1/random) to represent using valid enrollment speech for `2-speaker' condition 
and “0” “1” or “random” vectors for a pseudo enrollment representation in `1-speaker+noise' condition.

The `Mixture' row shows the metrics (SI-SNR, PESQ, and STOI) for the raw, unprocessed audio mixture. 
The SEF-PNet row corresponds to our previously proposed framework, serving here as a baseline. 
Notably, for the `1-speaker+noise' condition, we remove the ISA module from SEF-PNet, 
creating a purely SE-focused variant that relies only on the noisy input. 
This ensures a fair comparison since PSE functionality is unnecessary in a single-speaker context.
In contrast, the USEF-PNet row illustrates results from our proposed unified model 
that handles both PSE (2-speaker) and SE (1-speaker+noise) within a single architecture. 
When enrollment speech is provided, USEF-PNet effectively extracts the target speaker, 
while in the `1-speaker+noise' setting, it defaults to a conventional SE approach. 
Furthermore, the placeholders (0, 1, random) in the `1-speaker+noise' scenario evaluate how well 
the model manages a “dummy” enrollment representation.

By comparing the proposed USEF-PNet with its SEF-PNet baseline, we observe that USEF-PNet 
consistently outperforms or closely matches the baseline across both SE and PSE tasks, 
highlighting its strong generalization capabilities. Specifically, on the `2-speaker' test set, 
USEF-PNet improves SISDR/PESQ/STOI from 10.9/2.64/87.04 to 11.80/2.86/88.10, respectively, 
while on the `1-speaker+noise' test set, it achieves performance comparable to SEF-PNet. 
Furthermore, the results show that the (Y, 0) configuration yields the best SE performance, 
achieving a SISDR of 13.85, a PESQ of 2.95, and a STOI of 91.67. This suggests that a 
zero-valued enrollment speech avoids introducing extraneous interference, whereas both a 
constant “1” and random enrollment representation can slightly degrade SE performance. 
These findings confirm the robustness of USEF-PNet across both PSE and SE conditions, 
underscoring its practicality in real-world scenarios where enrollment speech may or 
may not be available. Moreover, this capability makes USEF-PNet particularly valuable 
for industrial applications, offering significant resource-saving benefits during deployment.

\subsubsection{Results on the full Libri2Mix `2-speaker' condition, and VoiceBank-DEMAND datasets}

In Table \ref{tab:Unvi-libri2mixvctk}, we present the performance of USEF-PNet on the full Libri2Mix `2-speaker' 
and VoiceBank-Demand datasets. To highlight the generalization capability of USEF-PNet in handling both 
SE and PSE tasks simultaneously, we replaced the `1-speaker+noise' SE task from Table \ref{tab:Unvi-libri2mix} 
with the VoiceBank-Demand dataset, a more commonly used dataset for conventional SE tasks, 
allowing for better comparison with state-of-the-art models. Additionally, for the `2-speaker' PSE condition, 
we utilize the full Libri2Mix dataset, which is more representative and comparable to public results.

In the `2-speaker' PSE condition, we compare USEF-PNet with sDPCCN and SEF-PNet. 
SEF-PNet, as the previous state-of-the-art model, achieves SISNR, PESQ, STOI of 13.00, 3.01, 89.71, respectively. 
However, USEF-PNet achieves SISNR, PESQ, STOI of 13.00, 2.99 and 89.74, demonstrating comparable performance to SEF-PNet.This confirms the effectiveness of USEF-PNet in PSE tasks without compromising overall performance.

For the SE task on the VoiceBank-Demand dataset, USEF-PNet is compared with two state-of-the-art models: 
MP-SENet and SEF-PNet. In this setting, USEF-PNet significantly outperforms SEF-PNet, 
improving  SISNR/PESQ from 19.21/3.54 to 19.69/3.59, while maintaining a similar STOI. Moreover, USEF-PNet also achieves competitive results with MP-SENet, further demonstrating the universality ability of the proposed USEF-PNet to simultaneously address both SE and PSE tasks.

\subsection{Results for Unsupervised Speech Disentanglement}

\begin{table}[t]  
\vspace{-1.5em}  
\renewcommand\arraystretch{1.3}
\caption{Condition-wise PSE results of DSEF-PNet on Libri2Mix.}
\label{tab:dsef3pse}
\centering
\scalebox{0.90}{
\begin{tabular}{l|l|c|c|c} 
\toprule
\textbf{Condition} & \textbf{Method} & \textbf{SISDR} & \textbf{PESQ} & \textbf{STOI} \\
\midrule
\multirow{3}{*}{1-speaker+noise} & Mixture  & 3.27 & 1.75 & 79.51 \\
                          & SEF-PNet   & 14.50 & 3.05 & 92.47 \\
                          & DSEF-PNet & \textbf{14.50} & \textbf{3.06} & \textbf{92.61} \\
\midrule
\multirow{3}{*}{2-speaker} & Mixture  & -0.03 & 1.60 & 71.38 \\ 
                          & SEF-PNet   & 13.00 & 3.01 & 89.71 \\
                          & DSEF-PNet & \textbf{13.57} & \textbf{3.08} & \textbf{90.65} \\
\midrule
\multirow{3}{*}{2-speaker+noise} & Mixture  & -2.03 & 1.43 & 64.65 \\ 
                          & SEF-PNet   & 7.54   & 2.14  & 80.58 \\
                          & DSEF-PNet & \textbf{7.57}  & \textbf{2.18}  & \textbf{80.66} \\
\bottomrule
\end{tabular}}
\end{table}

Table \ref{tab:dsef3pse} presents the condition-wise PSE results of SEF-PNet and DSEF-PNet on the Libri2Mix 
dataset under three different conditions: `1-speaker+noise', `2-speaker', and `2-speaker+noise'. 
The SEF-PNet serves as the baseline system, while DSEF-PNet represents our proposed model, 
which incorporates our unsupervised speech disentanglement strategy (the Heterogeneous 
Enrollment Invariant Training (HEIT) mechanism) on SEF-PNet.

Under the `1-speaker+noise' condition, SEF-PNet achieves SISDR/PESQ/STOI of 14.50/3.05/92.47, 
while DSEF-PNet slightly improves PESQ and STOI to 3.06 and 92.61, respectively, 
with SISDR remaining unchanged. These results suggest that DSEF-PNet refines speech quality and 
intelligibility while preserving the overall enhancement performance.
This suggests that in the single-speaker scenario, where both the enrollment speech 
and noisy speech correspond to the same speaker, our previous SEF-PNet is already 
capable of handling the task effectively. In the `2-speaker' condition, 
SEF-PNet achieves SISDR/PESQ/STOI = 13.00/3.01/89.71. 
DSEF-PNet improves these metrics to 13.57/3.08/90.65, indicating enhanced separation 
of the target speaker from interfering speech, leading to better overall speech quality 
and intelligibility. In this condition, HEIT plays a key role by guiding DSEF-PNet to 
focus on identity-related features 
while discarding irrelevant acoustic variations in the enrollment speech. 
This mechanism is especially beneficial in multi-speaker scenarios, as it helps the model 
isolate the target speaker more robustly, thereby boosting performance compared to SEF-PNet.

For the most challenging `2-speaker+noise' condition, SEF-PNet achieves SISDR/PESQ/STOI = 7.54/2.14/80.58, 
while DSEF-PNet slightly improves PESQ and STOI to 2.18 and 80.66, respectively, with a marginal increase in 
SISDR to 7.57. These results suggest that the additional noise complicates the disentanglement 
process of HEIT. However, by ensuring enrollment speech representations remain focused on the target 
speaker’s identity, HEIT helps the model maintain or slightly enhance its performance under these 
challenging acoustic conditions.

Overall, DSEF-PNet consistently outperforms our previous SEF-PNet across different conditions, 
with the most significant improvements observed in the `2-speaker' scenario. These findings 
highlight the advantages of DSEF-PNet in handling both single-speaker and 
multi-speaker scenarios while maintaining strong speech enhancement performance 
in noisy conditions.

\subsection{Results for USEF-PNet and DSEF-PNet combination}
\label{subsec:udsefpnet}

\begin{table}[!htbp] 
\renewcommand\arraystretch{1.3}
\caption{UDSEF-PNet on the full libri2mix 2-speaker condition and VoiceBank-Demand dataset.}
\label{tab:udsefpnet}
\centering
\scalebox{1.0}{
    \begin{tabular}{l|ccc|ccc}
    \toprule
    \multirow{2}{*}{\textbf{Methods}} 
    & \multicolumn{3}{c|}{\textbf{2-speaker}} 
    & \multicolumn{3}{c}{\textbf{VoiceBank-Demand}} \\
    \cmidrule(lr){2-4} \cmidrule(lr){5-7}
    & \textbf{SISDR} & \textbf{PESQ} & \textbf{STOI} 
    & \textbf{SISDR} & \textbf{PESQ} & \textbf{STOI} \\
    \midrule
    USEF-PNet & 13.00 & 2.99 & 89.74 & 19.69 & 3.59 & 94.26 \\
    UDSEF-PNet & \textbf{13.34} & \textbf{3.05} & \textbf{90.11} & \textbf{19.95} & \textbf{3.63} & \textbf{94.89} \\
    \bottomrule
    \end{tabular}}
\end{table}

After evaluating the effectiveness of USEF-PNet and DSEF-PNet individually, we combine both 
approaches into SEF-PNet to create UDSEF-PNet, which features a unified architecture for 
both SE and PSE tasks, incorporating the proposed HEIT for enrollment speech 
disentanglement. Table \ref{tab:udsefpnet} presents the results of UDSEF-PNet on the 
Libri2Mix `2-speaker' and VoiceBank-DEMAND datasets. 
In the `2-speaker' condition, UDSEF-PNet improves performance metrics from 13.00/2.99/89.74 
(USEF-PNet) to 13.34/3.05/90.11, demonstrating better speech separation and intelligibility. 
On the VoiceBank-DEMAND dataset, UDSEF-PNet further enhances performance, increasing 
SISNR/PESQ/STOI from 19.69/3.59/94.26 to 19.95/3.63/94.89. 
These results confirm that UDSEF-PNet's superior generalization and performance, 
making it a good candidate for real-world applications where both personalized speech 
enhancement and conventional speech enhancement are required.

\begin{table*}[t] 
\renewcommand\arraystretch{1.2}
\caption{Enrollment Speech Durations Examination Results on Libri2Mix `2-speaker' PSE condition.}
\label{tab:enrollment-dur}
\centering
\scalebox{1.0}{
    \begin{tabular}{l|c|ccc|ccc|ccc}
    \toprule
    \multirow{2}{*}{\textbf{Methods}} 
    & \multirow{2}{*}{\textbf{D$_e$}-$Train$}
    & \multicolumn{9}{c}{\textbf{D$_e$}-$Test$} \\
    & & \multicolumn{3}{c}{Random} 
    & \multicolumn{3}{c}{\textbf{$0-2s$}} 
    & \multicolumn{3}{c}{\textbf{$10-40s$}} \\
    \cmidrule(lr){3-5} \cmidrule(lr){6-8} \cmidrule(lr){9-11}
    & & \textbf{SISDR} & \textbf{PESQ} & \textbf{STOI} 
    & \textbf{SISDR} & \textbf{PESQ} & \textbf{STOI}
    & \textbf{SISDR} & \textbf{PESQ} & \textbf{STOI} \\
    \midrule
    \multirow{3}{*}{\textbf{SEF-PNet}} 
    & random 
    & 13.00 & 3.01 & 89.71 & 12.34 & 2.94 & 88.52 & 13.63 & 3.05 & 90.79 \\
    & \textbf{$2s$} 
    & 12.65 & 2.94 & 89.29 & 12.35 & 2.91 & 88.87 & 13.01 & 2.96 & 89.90 \\
    & \textbf{$10-40s$}
    & 12.95 & 2.97 & 89.66 & 12.41 & 2.92 & 88.63 & 13.57 & 3.02 & 90.75 \\
    \midrule
    \multirow{3}{*}{\textbf{DSEF-PNet}}
    & (random, random)
    & 13.57 & 3.08 & 90.65 & 13.05 & 3.05 & 89.84 & 13.98 & 3.12 & 91.34 \\
    & \textbf{($2s$, $10-40s$)}
    & 13.09 & 3.02 & 89.92 & 12.90 & 2.99 & 89.67 & 13.49 & 3.04 & 90.62 \\
    & \textbf{($10-40s$, $10-40s$)}
    & 13.16 & 3.03 & 89.91 & 12.69 & 2.98 & 89.20 & 13.66 & 3.07 & 90.86 \\
    \bottomrule
    \end{tabular}}
\end{table*}

\subsection{Examination of Long-Short Enrollment Pairing}

In Table \ref{tab:enrollment-dur}, we present an evaluation of SEF-PNet and DSEF-PNet under 
various enrollment speech durations, short ($0-2s$), long ($10-40s$) and random  (randomly selected from the training set), 
for the `2-speaker' condition in Libri2Mix PSE task. We also explore the long-short enrollment 
pairing (LSEP) strategy, which pairs short and long enrollment speech during training to enhance 
robustness.  \textbf{$D_{e}$}-\textit{Train} and \textbf{$D_{e}$}-\textit{Test} 
represent the enrollment speech duration used in the training and test stage, respectively, 
In DSEF-PNet, $(*, *)$ denotes the duration of two distinct enrollment speech recordings used 
for model training.

From Table \ref{tab:enrollment-dur}, we first observe that when both the training and test 
enrollment speech durations are short, both SEF-PNet and DSEF-PNet perform worse compared 
to scenarios where longer enrollment speech is used. This highlights the challenge of 
working with very short enrollment speech during both training and testing, as the model 
struggles to extract sufficient stable speaker-identity information in such limited-duration speech.
However, as we expected, both models tend to perform better with longer enrollment 
speech during inference test stage, indicating that more stable speaker information 
can be obtained from longer enrollment speech, thus 
yields better enhancement performance.

Regarding the impact of the LSEP strategy, although it is designed to enforce speaker identity consistency by aligning short and long enrollment representations, DSEF-PNet trained with one short and one long enrollment sample performs slightly worse than when trained with random enrollment durations. This may be attributed to the limited information in short enrollment speech, where forced alignment could suppress its discriminative power, and the model's potential reliance on long enrollment during training due to the absence of explicit modeling of duration differences.
Training with two long enrollment samples also yields lower performance than using random durations, which may be due to reduced variability in duration combinations during training and limited exposure to short enrollment conditions.

Interestingly, we also see from these results that long enrollment speech during training does 
not always guarantee the best results. Training with random durations often leads to stronger 
generalization across short, random, and long enrollment test conditions, 
suggesting that varied enrollment lengths help the model adapt to diverse scenarios. 
Finally, DSEF-PNet, which integrates unsupervised speech disentanglement, consistently 
outperforms SEF-PNet in all conditions, confirming the effectiveness of disentanglement-based 
strategies for robust personalized speech enhancement.

\section{Conclusion}

In this paper, we proposed two novel methods, USEF-PNet, DSEF-PNet, designed within a speaker embedding/ encoder-free unified framework for both 
Speech Enhancement (SE) and Personalized Speech Enhancement 
(PSE). USEF-PNet introduces a unified enrollment architecture that seamlessly integrates 
SE and PSE tasks into a single model, thereby simplifying deployment without requiring 
separate task-specific models. Building on this, DSEF-PNet incorporates an 
unsupervised enrollment speech disentanglement strategy that effectively separates 
speaker identity from irrelevant attributes, enhancing the robustness of the model. 
Meanwhile, LSEP strategy is explored to address scenarios with short 
enrollment speech by pairing it with long enrollment speech 
during training. While LSEP did not yield consistent improvements in our experiments, it offers a valuable perspective on designing training strategies for short enrollment speech scenarios.

Importantly, our proposed methods enhance performance on both conventional 
SE and PSE tasks compared to our previous SEF-PNet baseline, 
all while maintaining the original SEF-PNet architecture without increasing 
any additional parameters or model complexity. Extensive experiments on the 
Libri2Mix and VoiceBank-DEMAND datasets validate the effectiveness of our proposed approaches, 
demonstrating consistent improvements across various conditions. 
Our future work will focus on extending these methods to even more complex acoustic environments.

\section*{ACKNOWLEDGMENTS}
The work is supported by the National Natural ScienceFoundation of China (Grant No.62071302).

\bibliographystyle{IEEEtran} 
\bibliography{refs.bib}          

\end{document}